\documentclass[aps,prb,twocolumn,groupedaddress]{revtex4}

\usepackage[dvips]{graphicx}
\usepackage[dvips]{color}
\usepackage{amsmath}

\renewcommand{\Vec}[1]{\mbox{\boldmath$#1$}}

\begin{document}

\title{Two parameter flow of $\sigma_{xx}(\omega) - \sigma_{xy}(\omega) $ for the graphene quantum Hall system in ac regime}

\author{Takahiro Morimoto}
\affiliation{Department of Physics, University of Tokyo, Hongo, 
Tokyo 113-0033, Japan}
\author{Hideo Aoki}
\affiliation{Department of Physics, University of Tokyo, Hongo, 
Tokyo 113-0033, Japan}

\date{\today}

\begin{abstract}
Flow diagram of $(\sigma_{xx}, \sigma_{xy})$ in finite-frequency 
($\omega$) regime 
is numerically studied for graphene quantum Hall effect (QHE) system.  
The ac flow diagrams 
turn out to show qualitatively  similar behavior as the dc flow diagrams,
which can be understood that the dynamical length scale determined by the frequency 
poses a relevant cutoff for the renormalization flow.
Then the two parameter flow is discussed in terms of the dynamical scaling theory.
We also discuss the larger-$\omega$ regime which exhibits classical
flows driven by the raw frequency $\omega$.
\end{abstract}


\maketitle

\section{Introduction} 
In the quantum Hall effect (QHE), one standard and graphically clear way to grasp 
the physics involving the localization effect is the $\sigma_{xx} - \sigma_{xy}$ diagram, in which we look at the scaling flow (trajectories when the sample size is varied) of the longitudinal conductivity and Hall conductivity $(\sigma_{xx}, \sigma_{xy})$. 
The scaling property of the static QHE system, 
especially the quantization of the Hall conductivity into the multiples of $e^2/h$  and vanishing longitudinal conductivity, is beautifully captured with 
the $\sigma_{xx}-\sigma_{xy}$ diagram as originally discussed by 
Pruisken and  Khmelnitskii in terms of the non-linear sigma model\cite{pruisken-flow,pruisken-instanton,khmelnitskii}.  
For the conventional two-dimensional electron 
gas (2DEG), there exist 
(i) stable fixed points at $(\sigma_{xx},\sigma_{xy})=(0,n) (n$: integer), 
along with (ii) unstable fixed points characterizing delocalization at 
$(\sigma_{xx},\sigma_{xy})=(\sigma_{xx}^c,n+1/2)$.  
The attraction into the former, quantum-Hall fixed point accounts for the Hall insulating state with quantized values of Hall conductivity,
while the latter, unstable fixed point accounts for delocalized states at the center of each Landau level (LL), and dominates the behavior of the 
plateau-to-plateau transition in $\sigma_{xy}$.

Scaling properties of the Anderson transition have attracted both theoretical and experimental interests, 
since they should be universal and depend only on the symmetry class of the system \cite{huckestein}.
The critical exponent has been numerically studied for the lowest LL \cite{huckestein-kramer}, and 
later with the Chalker-Coddington network model \cite{slevin-ohtsuki} and experimentally confirmed by Li et al. \cite{Li-scaling-exp} 
The universal value of longitudinal conductances at the LL centers are intensively discussed with a tight binding lattice model \cite{schweitzer-markos} .
Thus the scaling behavior at the plateau-to-plateau transition has been established.

On the other hand, rapid advances in the terahertz (THz) spectroscopy technique 
have made the optical responses of the quantum Hall system, such as cyclotron resonances and Faraday rotations, 
experimentally accessible\cite{hangyo,ikebe2008cds}.
Specifically, the Faraday rotation is proportional to the optical Hall conductivity $\sigma_{xy}(\omega)$, and we have an intriguing problem 
of how the static 
Hall conductivity, which may be regarded as a topological quantity\cite{tknn}, evolves into the optical Hall conductivity, especially in the relevant 
(cyclotron) energy scale which falls upon the THz regime. \cite{mikhailov85,gusynin-sxy,morimoto-opthall,Fialkovsky09}
On the optical Hall conductivity, we have recently shown, numerically, 
that the plateau structure in $\sigma_{xy}(\omega)$  is unexpectedly 
retained in the ac (THz) regime in both 2DEG and in graphene, 
although the plateau height deviates from the quantized 
values in ac.\cite{morimoto-opthall}.   
Graphene is particularly interesting, since 
a massless Dirac system is realized as the low-energy physics, 
which accommodates a novel Dirac QHE is observed \cite{Nov05,Kim-gr}, 
for which the scaling theory of QHE in graphene has been formulated in terms of the non-linear sigma model \cite{ostrovsky08}.
 Experimentally, the ac plateau has been observed 
in a GHz Faraday rotation measurement for a 
2DEG system \cite{hohls-kuchar02} and recently in THz regime \cite{ikebe-THz}.  
In the graphene QHE system, we expect plateau structures at the tail 
(small frequency) region, while 
Faraday rotation was measured in the region around cyclotron resonances \cite{crassee2010giant}.

Considering these advances in spectroscopies of QHE systems,
it is important to study the systematic behavior of plateau structure in ac and optical regimes.
Robustness of the ac plateau structure against disorder 
as revealed in the numerical result can be understood if 
we consider the effect of localization, which dominates the physics of electrons 
around the centers of Landau levels in disordered QHE systems, 
in finite-frequency regime, following the scaling theory of the Anderson transition.  
Namely, a finite frequency put an effective cutoff (the dynamical length scale $L_\omega$)
 for the system, and the plateau in the ac Hall conductivity should be retained in the region where  the localization length ($\xi \sim |\epsilon-\epsilon_c|^{-\nu}$)  diverging toward LL center is smaller than the dynamically posed cutoff  $L_\omega$.
In the two parameter flow picture, this can be viewed 
as the dynamical length scale $L_\omega$ determining a scale where the renormalization is stopped.

The dynamical scaling behavior has been studied for the ac longitudinal conductivity \cite{gammel-brenig},
and for the optical Hall 
$\sigma_{xy}(\omega)$ \cite{morimoto-ac-scaling}.
Now, it is interesting to combine  
both the optical longitudinal and Hall $\sigma_{xy}(\omega)$ 
and numerically map out the two parameter ($\sigma_{xx} - \sigma_{xy}$) flow diagram 
in the ac regime.  

In the present work we have calculated  both optical longitudinal and Hall conductivities ($\sigma_{xx}, \sigma_{xy}$ ) for the graphene QHE system with a potential disorder,
 and combined them to 
numerically examine 
the two parameter $\sigma_{xx} - \sigma_{xy}$  flow diagram in the ac regime.
We study $n=0$ LL in the graphene QHE system with exact diagonalization method to treat disorder effects.  
 One particular point of interest 
is the behavior around fixed points in the $\sigma_{xx}(\omega) - \sigma_{xy}(\omega)$ diagram.  
There, we have focused on the $n=0$ Dirac Landau level, 
where the peculiarity of graphene appears as the 
property that $n=0$ is an electron-hole symmetric point. 
In a small-$\omega$ regime,
we obtain numerical results which are coherent with the above picture that 
the $\sigma_{xx}(\omega) - \sigma_{xy}(\omega)$ flow
obeys the Pruisken's two-parameter flow,
with $L_\omega$ as a relevant cutoff for the system in ac region, 
 where the flows are between 
$\sigma_{xy} = \pm 2 e^2/h$ reflecting the graphene QHE including the valley and the spin degeneracies.
We also discuss a large-$\omega$ regime where the frequency $\omega$ is comparable with the cyclotron frequency $\omega_c$
and  exhibits classical flows driven by the raw frequency.

\section{Formalism}  
For graphene QHE system, we employ 
the two-dimensional effective Dirac model,
\begin{equation}
H=v_F \tau_z {\boldsymbol \sigma} \cdot {\boldsymbol \pi} + V(\Vec{r}),
\end{equation}
where $v_F$ is Fermi velocity, 
${\boldsymbol \sigma}=(\sigma_x,\sigma_y)$ and $\tau_z$ the 
Pauli matrices acting on the space of two sublattices (A, B) and  two valleys (K, K'), ${\boldsymbol \pi}=\Vec{p}+e\Vec{A}$ with 
 $\Vec{p}=(p_x,p_y)$ the momentum, and $\Vec{A}$ the vector potential. 
Disorder is introduced by a random potential, 
$$
V(\Vec{r})=\sum_{i,j} u_{i,j} \exp(-|\Vec r- \Vec{R}_{i,j}|^2/2d^2)/(2\pi d^2), 
$$
composed of 
Gaussian scattering centers of range $d$ and 
$u_{i,j}$ takes a value in $(-u,u)$ randomly. 
Here we take $d=0.7\ell$, where $\ell=\sqrt{\hbar/eB}$ is the magnetic length.  For numerical facility, impurity sites $R_{i,j}$ are periodically placed on
$
R_{i,j}= (2\pi\ell^2/L)(i,j)
$ with $L$ being the linear dimension of the sample. 
A measure of disorder is given by the Landau level broadening\cite{ando}, 
$
\Gamma = 2 u [N_{imp}/2\pi(\ell^2+2d^2) L^2]^{1/2},
$
with a number of impurity sites $N_{imp}$.
We assume smooth potential disorders in the length scale of underlying lattice structure,
and we neglect inter-valley scattering. 
The cyclotron energy is, for a Dirac particle,  given by 
$\omega_c = \sqrt{2}v_F/\ell$.

Since we are interested in the dynamical $\sigma_{xx}(\omega) - \sigma_{xy}(\omega)$, which should be related to the localization physics, 
we obtain the eigenstates of the Hamiltonian with an exact 
diagonalization, which is done for a subspace 
spanned by a finite number of Landau levels (LL's) around $n=0$ LL,
for $L\times L$ systems with $L/\ell$ varied over $20, 30, 40$.
Here we retain 5 LLs ($n= -2 \sim 2$), which poses an ultraviolet cutoff.
\footnote{
For each valley, this choice of high-energy cutoff (retaining $n=-N_{max} \sim N_{max}$ LLs) makes the Hall conductivity 
coincide with the half of the total Hall conductivity contributed from both two valleys (K, K'),
for which a cancellation of a ultraviolet divergence occurs \cite{ostrovsky08}.
So we can concentrate on one of the decoupled valleys in the numerical calculation,
although well-defined conductivities are the sum of contributions from two valleys
}
In the Landau gauge $\Vec{A}=(0, Bx)$, 
the basis function is $\psi_{n,k}=e^{-iky} \phi_n(x-\ell^2 k_y)$,
where $\phi_n$ is the Dirac-Landau function in the $n$-th 
Landau level \cite{ZhengAndo},
and wavenumbers k takes an integer multiples of $2\pi/L$ with a periodic boundary condition for $y$-direction.
The number of discrete 
wavenumbers $N_k$ is related to $L$ with $N_k=L^2/2\pi \ell^2$ in 
a finite system.  From 
the eigenfunctions $\psi_a$ and eigenenergies $\epsilon_a$ obtained 
with the exact diagonalization, 
the optical Hall conductivity\cite{morimoto-opthall} is evaluated from the Kubo formula \cite{kubo1965} as 
\begin{equation}
\sigma_{xy}(\omega) =
 \frac{\hbar}{iL^2} \sum_{ab} j_x^{ab} j_y^{ba} 
\frac{f(\epsilon_b) - f(\epsilon_a)}{\epsilon_b-\epsilon_a}
\frac{1}{\epsilon_b-\epsilon_a-\hbar\omega-i\eta},
\end{equation}
where $f(\varepsilon)$ is the Fermi distribution, and 
$\eta$ a low-energy cutoff.  The current matrix 
element, $j_x^{ab}$, has a selection rule peculiar to 
Dirac model 
($n \leftrightarrow \pm n \pm 1$ with $n$ the Landau index), which 
is  distinct
from that ($n \leftrightarrow n\pm 1$) for 2DEG as
\begin{eqnarray*}
j_{x}^{n,n'}=e v_F C_{n} C_{n'}
\left[{\rm sgn}(n)\delta_{|n|-1,|n'|}+{\rm sgn}(n')\delta_{|n|+1,|n'|}\right],\\
j_{y} ^{n,n'}=i e v_F C_{n} C_{n'}\left[{\rm sgn}(n)\delta_{|n|-1,|n'|}-{\rm sgn}(n')\delta_{|n|+1,|n'|}\right],
\label{matrixelement}
\end{eqnarray*}
where $C_n= 1 (n=0)$ or $1/\sqrt{2}$ (otherwise) \cite{shon-ando,ZhengAndo}.

The longitudinal conductivity, on the other hand, is given by 
$$
\mbox{Re} \sigma_{xx}(\omega) 
= 
\frac{\hbar}{L^2}
\sum_{\varepsilon_a, \varepsilon_b}
\frac{f(\varepsilon_b)-f(\varepsilon_a)}{\varepsilon_b-\varepsilon_a} 
\frac{|j_{x}^{ab}|^2 \eta}{(\varepsilon_b-\varepsilon_a-\hbar \omega)^2+\eta^2}.$$
We note that the low-energy cutoff  $\eta$, which 
affects the $\omega \sim 0$ behavior of $\sigma_{xx}(\omega)$, 
should be chosen close to the Thouless energy, which is 
typically of the order of the energy level spacing
$\sim 1/L^2$.\cite{thouless-kirkpatrick,nomura-ryu-koshino,nomura2008qhe}
The temperature in the Fermi distribution function $f(\epsilon)$ is 
set to be small as far as the low-frequency behavior of $\sigma_{xx}(\omega)$ is numerically stable, which is achieved if we 
put the temperature to be of the order of the level spacing $\sim 1/L^2$.
 For the scaling analysis 
the calculation is repeated for varied sample size $L$, Fermi energy $\varepsilon_F$ and 
 frequency $\omega$.  Throughout the paper 
the length, energy and frequency are respectively in units of $\ell, \hbar\omega_c$ and $\omega_c$.

\section{$\sigma_{xx}-\sigma_{xy}$ diagram in ac regime}
\begin{figure}[tb]
\begin{center}
\includegraphics[width=0.7\linewidth]{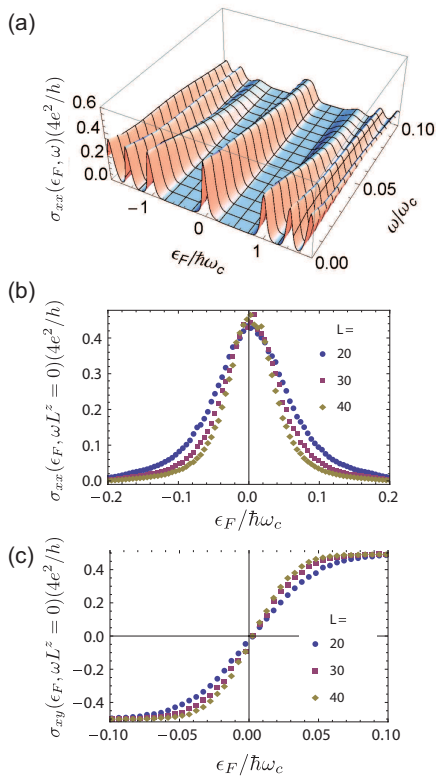}
\end{center}
\caption{
(a) $\sigma_{xx}(\epsilon_F,\omega)$ plotted 
against the Fermi energy $\epsilon_F$ and the frequency $\omega$
for $L=30$.  Lower panels depict 
$\sigma_{xx}(\omega=0)$ (b) and $\sigma_{xy}(\omega=0)$ (c) 
for various sample sizes 
with disorder strength $\Gamma/\hbar \omega_c=0.4$.
}
\label{bare-sigma}
\end{figure}

First we discuss the behavior $\sigma_{xx}(\omega)$ and $\sigma_{xy}(\omega)$ separately.
$\sigma_{xx}(\epsilon_F, \omega)$  in Fig.\ref{bare-sigma}(a) 
shows a ridge structure along the $\omega$ axis 
when the Fermi energy $\epsilon_F$ is around the 
delocalized region at each LL.  
When we increase $L$ in Fig.\ref{bare-sigma}(b), 
the width of $\sigma_{xx}(\omega=0)$ becomes narrower, 
while the peak height stays almost constant
which is expected from the universal longitudinal conductance\cite{kivelson-global-phase-daigram,schweitzer-markos,wong-sxxc}.  
On the other hand, 
$\sigma_{xy}(\epsilon_F, \omega)$ plotted against $\epsilon_F$ 
in Fig.\ref{bare-sigma}(c) shows a transition 
from $\sigma_{xy}=-2$ plateau to  $\sigma_{xy}=2$ plateau 
around $n=0$ LL with 2 valley and 2 spin degeneracies, where 
the transition width of $\sigma_{xy}(\omega=0)$ sharpens with 
increasing $L$.

\begin{figure}[tb]
\begin{center}
\includegraphics[width=0.95\linewidth]{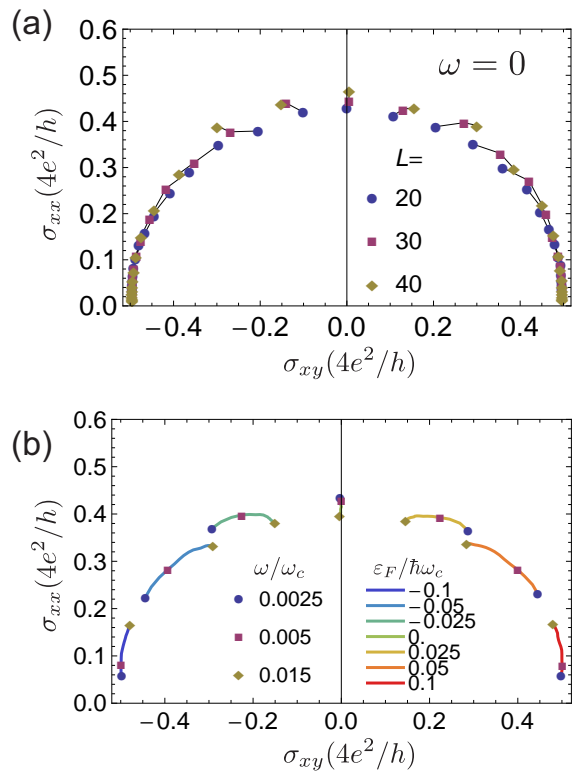}
\end{center}

\caption{
The flow of $(\sigma_{xx}(\omega)-\sigma_{xy}(\omega))$ in the graphene quantum Hall system  with a disorder strength $\Gamma/\hbar\omega_c=0.4$.
(a) The flow in dc regime ($\omega= 0$) for various Fermi energy $\varepsilon_F$ and system size  $L/\ell=20,30,40$.
(b) The flow in ac regime for various values of Fermi energy $\varepsilon_F$ and frequency $\omega$ with a fixed system size $L/\ell=30$
}
\label{smallw}
\end{figure}

\begin{figure}[tb]
\begin{center}
\includegraphics[width=0.95\linewidth]{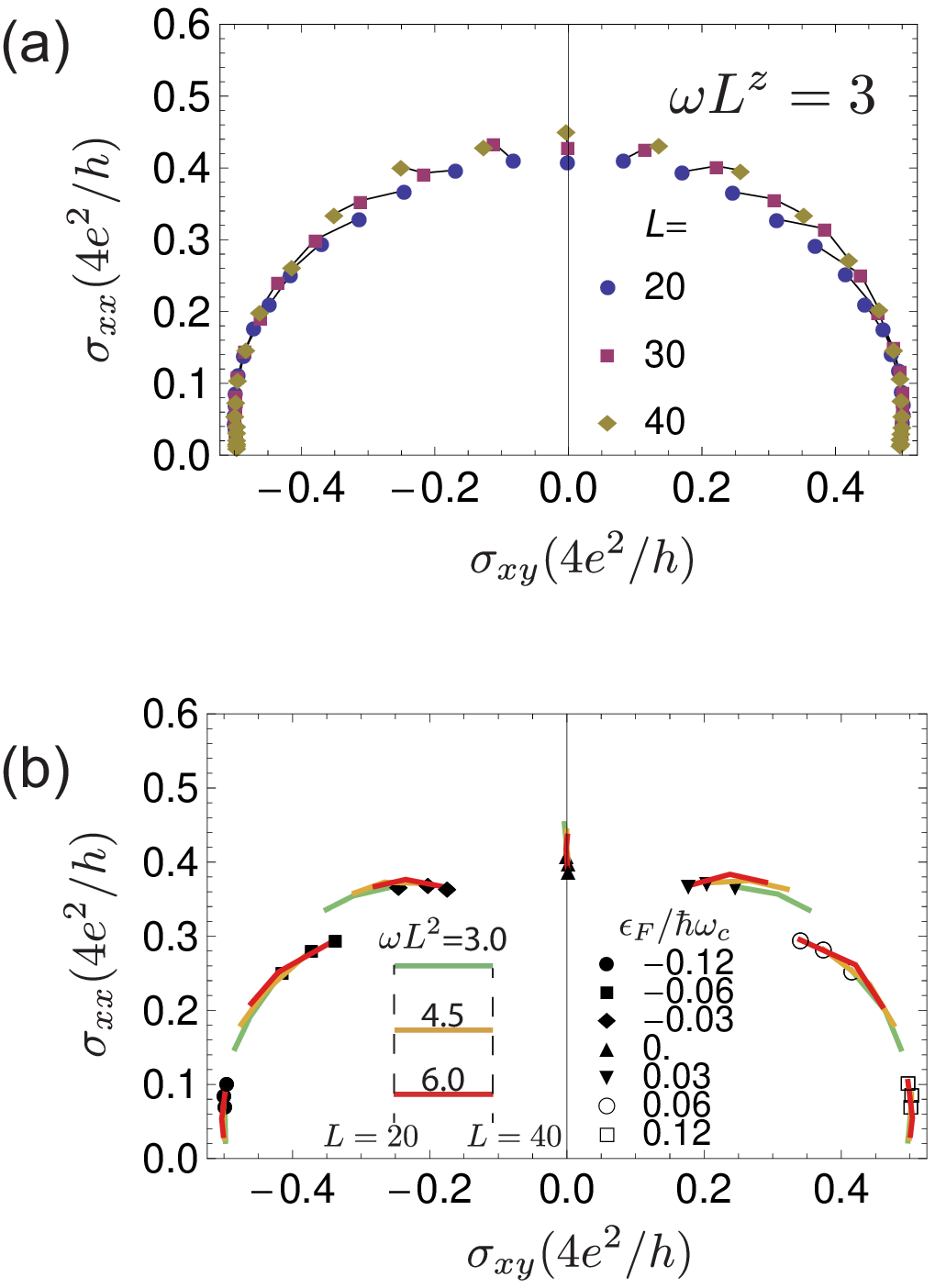}
\end{center}

\caption{
 (a) The flow of $(\sigma_{xx}(\omega)-\sigma_{xy}(\omega))$ in the graphene quantum Hall system
for renormalized frequency $\omega L^z= 3$
with $z=2$ and $L=20,30,40$ with a disorder strength $\Gamma/\hbar\omega_c=0.4$.  
(b) Flows when the sample size is varied as $L=20\rightarrow 40$ for a 
fixed of $\omega L^z$.   
For each value of  $\omega L^z$ we plot the flows corresponding 
to various values of Fermi energy $\epsilon_F$.  
The value of $\epsilon_F$ is indicated by different symbols (circles, squares, etc) 
that mark the smallest sample size ($L=20$). 
The results for various values (color-coded) of $\omega L^z=3-6$ are superposed.  
}
\label{flow-scaling}
\end{figure}

Now we are in a position to examine $\sigma_{xx}-\sigma_{xy}$ diagram in Fig.\ref{smallw}.
First, the diagram for $\omega=0$ is depicted in Fig.\ref{smallw}(a), 
where almost all the points are attracted to 
the points $(\sigma_{xx},\sigma_{xy})=(0, \pm 2)e^2/h$ with increased 
sample size $L$,
while the point sitting at 
$\sigma_{xy} = 0$ only exhibits a tiny, upward flow.  
This numerically calculated dc flow diagram is clearly understood in terms of Pruisken's two parameter flow picture \cite{pruisken-flow,pruisken-instanton,khmelnitskii}.
We can interpret the static result that  
the flows starting from $\sigma_{xy} \neq 0$ corresponding to 
those flowing into 
the stable fixed points at $(\sigma_{xx}, \sigma_{xy})=(0, \pm 2)e^2/h$ which describes Hall plateau for graphene,
while the tiny upward flow around $\sigma_{xy} = 0$ 
corresponds to the flow that starts from the point with a $\sigma_{xx}$ smaller than $\sigma_{xx}^c$ (rather than larger $\sigma_{xx}$ expected from SCBA values), 
 and shows a renormalization to the  unstable fixed point at 
$(\sigma_{xx}, \sigma_{xy})=(\sigma_{xx}^c, 0)$.
So we are seeing the region below the unstable fixed point.
This upward flow toward the unstable fixed point would reflect the existence of the delocalized state at the LL center,
since it is percolating through the sample and has a metallic nature. 
The longitudinal conductivity then increases with the sample size and converges to the universal conductance at the LL center.
This dc result for graphene $n=0$ LL is consistent with Nomura et al who 
discusses the Thouless-number and  the Hall conductivity 
 for the dc flow diagram.\cite{nomura2008qhe}

If we now turn to the ac result that is plotted for a fixed system size $L/\ell=30$ 
varying frequency $\omega/\omega_c=0.0025 \sim 0.015$ in 
Fig.\ref{smallw}(b), where a behavior quite similar to 
the dc data is found.  Namely, almost all the points away from the LL center
 are attracted to the QHE fixed points, 
while the point on $\sigma_{xy}= 0$ at the center of LL flows 
only slightly shift upwards.  
This behavior is understood  that, 
in this small frequency regime, 
the relevant cutoff length scale for the critical behavior of localization length $\xi$ or the renormalization equation for two-parameter flow is posed by the frequency through the dynamical length scale $L_\omega $ 
instead of the sample size $L$ in the dc regime,
so that the overall behavior is determined by the same two-parameter flow, where 
the effective cutoff alone is changed systematically with the frequency $\omega$ 
as $L_\omega \sim \omega^{-1/z}$ .

\section{Dynamical scaling analysis}
Now let us describe the dynamical scaling 
for $\sigma_{xx}(\varepsilon_F,\omega)$ and $\sigma_{xy}(\varepsilon_F,\omega)$.
The scaling argument starts from 
an ansatz that the optical conductivity depends on 
Fermi energy $\varepsilon_F$ and frequency $\omega$
only through the ratios $L/\xi$ and $L_\omega/\xi$.
The physical quantities should then be described in terms of 
the universal scaling function of the ratios $L/\xi$ and $L_\omega/\xi$.
Here $\xi$ is the localization length 
with a critical behavior $\xi \sim 1/|\varepsilon_F - \varepsilon_c|^\nu$, 
where $\varepsilon_c$ is the critical energy which 
coincides with the center of the LL ($\varepsilon_c=0$ for $n=0$), and 
$\nu$ the localization critical exponent.
The dynamical length scale, which 
is the distance over which an electron travels during one cycle, 
$1/\omega$, of the ac field, is assumed to 
behave as $L_\omega \sim 1/\omega^{1/z}$, 
where $z$ is the dynamical critical exponent, 
assuming $z= 2$ in this paper
since we treat the non-interacting electrons \cite{huckestein-z}, 
while $z=1$ is established for the case with electron-electron interaction \cite{shklovskii}.
For these critical behaviors the dynamical scaling ansatz 
for the 
longitudinal and transverse conductivities amounts to \cite{chalker-scaling-ansatz}
\begin{eqnarray}
\sigma_{xx}(\varepsilon_F,\omega,L)=\frac{e^2}{h} 
F_{xx}(\delta\varepsilon_F L^{1/\nu},\omega L^z),
\nonumber\\
\sigma_{xy}(\varepsilon_F,\omega,L)=\frac{e^2}{h} 
F_{xy}(\delta\varepsilon_F L^{1/\nu},\omega L^z),
\label{scaling-eq}
\end{eqnarray}
where $F_{xx},F_{xy}$ are universal scaling functions,
and $\delta\varepsilon_F \equiv \varepsilon_F-\varepsilon_c$ .

This ansatz is only valid for the critical region,
 where the deviation of Fermi energy from the LL center $\delta\varepsilon_F$ is assumed to be small and the frequency $\omega$ also to be small.
In Sec.3 and Sec.4, we precisely consider this region with small $\delta\varepsilon_F$ and $\omega$, 
while in Sec.5 where we shall treat a large-frequency region 
this ansatz should be no longer applicable.
The ansatz indicates that the flow of $(\sigma_{xx}(\omega),\sigma_{xy}(\omega))$ in the ac region 
 depends on the frequency $\omega$ only through the ratio of the  rescaled frequency with the system size as $\omega L^z$. 

We now interpret the behavior of $\sigma_{xx}-\sigma_{xy}$ diagram in terms of the dynamical scaling eqn.(\ref{scaling-eq}) as shown in Fig.\ref{flow-scaling}.
When we consider a dynamical scaling behavior,
it is convenient to consider a rescaled frequency $\omega L^z$ 
since a dependence on the frequency appears in  a form of $\omega L^z$ in eqn.(\ref{scaling-eq}).
In Fig. \ref{flow-scaling}(a) we show a result for a fixed rescaled frequency $\omega L^z=3$,
then the flow of ac conductivities can be discussed in terms of varying $L$ as in dc case.The dynamical scaling hypothesis (eqn.(\ref{scaling-eq})) expects that  
$(\sigma_{xx},\sigma_{xy})$ right on $\varepsilon_F=0$ with $\delta \epsilon_F L^{1/\nu} =0$ for all $L$ should depend only on $\omega L^z$,  i.e., does not flow, 
while at the center of Fig.\ref{flow-scaling}(a) a slight upward flow is seen showing a metallic behavior discussed above.  
Away from $\varepsilon_F = 0$, on the other hand, 
the flow should depend only on 
$\delta\varepsilon_F L^{1/\nu}$ from eq.(\ref{scaling-eq})
for a fixed value of the rescaled frequency $\omega L^z$, 
which implies that the flows starting from various $\varepsilon_F \neq 0$ 
should reside on a single curve as seen in Fig.\ref{flow-scaling}(a). 

In order to examine the dependence of the two parameter flow on the rescaled frequency $\omega L^z$,
 we superpose all the results 
in Fig.\ref{flow-scaling}(b).  
There, 
we show flows when the sample size is varied as $L=20\rightarrow 40$ for fixed values of $\omega L^z$.   
For each value of  $\omega L^z$ we plot the flows corresponding 
to various values of Fermi energy $\epsilon_F$.  
The value of $\epsilon_F$ is indicated by different symbols 
that mark the smallest sample size ($L=20$). 
Fig.\ref{flow-scaling}(b) thus visualizes the 
the two-parameter flows superposes for various values 
of $\omega L^2$ and  for various values  of Fermi energy $\epsilon_F$.
The results for various values of $\omega L^z=3-6$ are then superposed 
in Fig.\ref{flow-scaling}(b).  
In this summary plot we can see 
that the 
$(\sigma_{xx},\sigma_{xy})$  
flows for 
different values of  $\omega L^z$ tend to coalesce into a single curve
in the region away from $\sigma_{xy}=0$.
This is a consequence 
that we see the same two parameter flow with various cutoff length scale posed by different rescaled frequencies.
Close to the unstable fixed point at $(\sigma_{xx},\sigma_{xy}) = ( \sigma_{xx}^c ,0)$ the flow shows a metallic behavior (i.e., $\sigma_{xx}$ increasing with $L$) renormalizing into the unstable fixed point with increasing $\sigma_{xx}$, 
and slightly deviates from a single curve.   
The role of the value of increasing $\omega L^z$ appears in a shift of the initial position of the flow 
in $(\sigma_{xx}(\omega), \sigma_{xy}(\omega))$ 
for each $\varepsilon_F$ toward the opposite direction to the flow.
More precisely, for a larger $\omega L^z$ 
we have more broadened peak structure in $\sigma_{xx}$ and 
 more broadened transition width in $\sigma_{xy}$, so that 
the initial point (smallest-$L$ data) of the flow 
for each $\varepsilon_F$ shifts closer to the unstable fixed point  
(the center of the flow at $(\sigma_{xx}, \sigma_{xy}) = (\sigma_{xx}^c, 0)$ ), 
which we can observe in a shift of initial values in Fig.\ref{flow-scaling}(b).
This behavior arises from the fact that 
the relevant cutoff for the two-parameter flow in the ac regime is determined by the dynamical length scale $L_\omega$, 
where a larger frequency $\omega$ gives a smaller cutoff length scale $L_\omega \sim \omega^{-1/z}$  and leads to 
an overall shift toward the direction opposite to that of the flow 
 (in this case, toward the unstable fixed point), 
 and a more broadened width of the plateau-to-plateau transition.

\section{Flow diagrams for larger $\omega$}

\begin{figure}[tb]
\begin{center}
\includegraphics[width=0.9\linewidth]{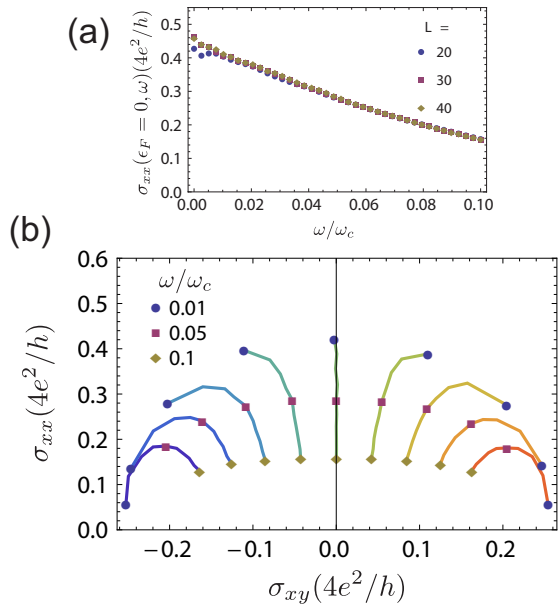}
\end{center}
\caption{
(a) $\sigma_{xx}(\epsilon_F=0,\omega)$ plotted for frequency up 
to $\omega=0.1$, 
where the data for various system sizes almost 
coincide with each other for large $\omega$.
(b) $\sigma_{xx}(\omega)-\sigma_{xy}(\omega)$ diagram 
for raw frequency $\omega$ with $L=30$.
}
\label{large-w}
\end{figure}

So far we have discussed the behavior for a small-frequency region, 
where the dynamical length scale $L_\omega \sim \omega^{-1/z}$ behaves an infrared cutoff  for the critical phenomena at the Anderson transition
and the dynamical scaling arguments also hold quite well.  
It is worth while to ask how  the behavior of the larger-frequency region,
typically for $\omega$ up to $0.1\omega_c$,
 looks like  in a similar two parameter $\sigma_{xx}-\sigma_{xy}$ plot. 
Naturally, in the large-frequency region, ac conductivities are expected to be no longer dominated by the  criticality
and  show a qualitatively different, rather classical behavior.
In this region we should adopt raw $\omega$ instead of 
$\omega L^z$, 
because the system should be out of the critical (i.e., dynamical-scaling) 
region. 
In Fig.\ref{large-w}(a), we look at the large-$\omega$ behavior of the 
longitudinal $\sigma_{xx}(\omega,\varepsilon_F=0)$ against 
the frequency $\omega$, 
from $\omega = 0.01\omega_c$  to  $\omega = 0.1\omega_c$.  
In a large-frequency, ac longitudinal conductivity at the center of LL shows a monotonic decrease of $\sigma_{xx}(\varepsilon=0)$ with $\omega$ 
as consistent with Ref.\cite{gammel-brenig}.  
This clearly signals a deviation from the dynamical scaling ansatz, which assumes the $\omega L^z$-dependent conductivity as in eqn.\ref{scaling-eq}.

The $\sigma_{xx}(\omega)-\sigma_{xy}(\omega)$ diagram 
for the large-$\omega$ region where the flows are 
indicated with varying frequency $\omega$ 
for various values of $\epsilon_F$ 
is shown in Fig.\ref{large-w}(b).
The $\omega$-driven flows show a different pattern from the flows in a small-omega region in the previous section,
although it is  attracted 
with decreasing $\omega$  to the static-QHE fixed points $(\sigma_{xx},\sigma_{xy})=(0, \pm 2)e^2/h$,  just as in the 
temperature-driven flows\cite{aoki1986-T-flow}.   
This behavior is understood 
that in the large-frequency region, where frequency $\omega$ become comparable to the cyclotron frequency $\omega_c$,
the frequency puts a small length scale comparable to the magnetic length, 
which should naturally induce deviations from the critical region and
show a pattern of the two conductivities different from that in the small-frequency region.
The existence itself of 
flows around $\omega/\omega_c \sim 1$  implies 
that the system is not fully dominated by a Drude-like behavior, 
which is consistent with the observation 
of the ac plateau structure in this frequency region, i.e., 
$\omega \sim 0.1$ that corresponds to the THz region.

\section{Summary}
We have numerically obtained $\sigma_{xx}(\omega)-\sigma_{xy}(\omega)$ diagram for graphene QHE system, and 
have examined  the flow for two regimes.
In a small-$\omega$ regime,
flows are governed by the dynamical length scale posed by the frequency
as a relevant cutoff length scale for the criticality around the Anderson transition.
We also discussed a metallic behavior around the unstable fixed point 
reflecting the delocalized state at the LL center.
The larger-$\omega$ regime exhibits 
rather classical flows driven by the bare frequency
due to the small dynamical length scale comparable to the magnetic length scale .

We wish to thank Mikito Koshino, Kentaro Nomura and Akira Furusaki for 
illuminating discussions.  
This work has been supported in part by Grants-in-Aid for Scientific Research, 
Nos.20340098, 23340112 from JSPS. TM has been supported by JSPS.


\end{document}